\documentclass[aps,pra,reprint,superscriptaddress]{revtex4-2}

\usepackage{graphicx} 
\usepackage{xcolor} 
\usepackage{siunitx}
\usepackage{ulem}  
\usepackage{amsmath}
\usepackage{amssymb}

\usepackage{todonotes}
\usepackage[dvipsnames]{xcolor}

\usepackage{xr-hyper}
\usepackage{hyperref}

\begin{document}

\title{Motion-induced directionality of collective emission in a non-chiral waveguide
}

\author{Yoan Spahn}
\affiliation{Institute of Applied Physics, \href{https://ror.org/05n911h24}{Technical University of Darmstadt}, Hochschulstra\ss e 6, 64289 Darmstadt, Germany}
\author{Jens Hartmann}
\affiliation{Dept. of Physics and research center OPTIMAS, \href{https://ror.org/01qrts582}{RPTU University Kaiserslautern-Landau}, 67663 Kaiserslautern, Germany}
\author{Benedikt Saalfrank}
\affiliation{Institute of Applied Physics, \href{https://ror.org/05n911h24}{Technical University of Darmstadt}, Hochschulstra\ss e 6, 64289 Darmstadt, Germany}
\author{Michael Fleischhauer}
\affiliation{Dept. of Physics and research center OPTIMAS, \href{https://ror.org/01qrts582}{RPTU University Kaiserslautern-Landau}, 67663 Kaiserslautern, Germany}
\author{Thomas Halfmann}
\affiliation{Institute of Applied Physics, \href{https://ror.org/05n911h24}{Technical University of Darmstadt}, Hochschulstra\ss e 6, 64289 Darmstadt, Germany}
\author{Thorsten Peters}
\affiliation{Institute of Applied Physics, \href{https://ror.org/05n911h24}{Technical University of Darmstadt}, Hochschulstra\ss e 6, 64289 Darmstadt, Germany}

\date{\today}

\begin{abstract}
We report the experimental observation of motion-induced directionality in collective atomic emission within a hollow-core waveguide, establishing a general principle: directional interactions can emerge from collective phase engineering alone. Remarkably, neither single-emitter asymmetry nor any asymmetry in the geometric arrangement of the system is required - both the atom-field coupling and the spontaneous emission are fully isotropic in our system. Instead, Raman-induced effective two-level emitters with spatially oscillating transition dipole phases and atomic motion give rise to controllable directionality, reaching values up to 0.89(1).
We study the correlations of the superfluorescent bursts close to and well above the threshold to collective emission; we find thermal statistics below and a buildup of coherence above it.
Numerical simulations based on the Truncated Wigner Approximation for spins yield good agreement. Additionally we present a simple model based on position uncertainty capable of reproducing the observed directionality.
Our results open a new route to directional interactions in non-chiral systems, with direct implications for the design of directional metamaterials and photonic structures built from isotropic constituents.

\end{abstract}

\maketitle


Understanding and controlling the coupling in light-matter interfaces is one of the main goals of quantum optics. A basic setting involves the emission of photons by an ensemble of emitters. 
In dilute ensembles the interaction of quantum emitters with vacuum fluctuations of the radiation field creates  oscillating atomic dipoles, which have a random phase leading to isotropic spontaneous emission of light and 
exponential decay.
This situation is quite different in dense \cite{D54,GH82,PhysRevA.3.1735,FGR21} or waveguide-coupled ensembles \cite{SGH17,PBJ22,LTB24,SBP24}. Here, dissipative dipole-dipole couplings between emitters mediated by exchange of photons become important. These synchronize the atomic dipoles and modify the emission process. As a consequence, collective scattering in form of superradiant light bursts \cite{D54,GH82} can emerge with a preferred bidirectional emission along the axis of largest optical depth of the ensemble and non-exponential decay. 
By coupling the emitters to a chiral interface \cite{LMS17,SJV25} as, e.g., realized by photonic nanostructures \cite{PVR14,IP21,MA25} or chiral arrangement of individual emitters \cite{POY24,XPB26}, even unidirectional emission can be achieved. These interfaces have seen an increased interest in recent years as they give rise to nonreciprocal interactions \cite{fruchart2021non,CAT18,KKR21,NCK22,NBB25} with applications for, e.g., photonic metamaterials \cite{GTY22}, nanophotonics \cite{SIG23,LGL22,yang2024nonreciprocal,RMX22}, magnonics \cite{CYG22}, and many-body systems \cite{BMV21}. For these interfaces unidirectional guiding occurs at the level of single emitters or due to their spatial arrangement and thus also affects collective emission \cite{LTB24}.

\begin{figure}[b]
\includegraphics[width=0.97\columnwidth]{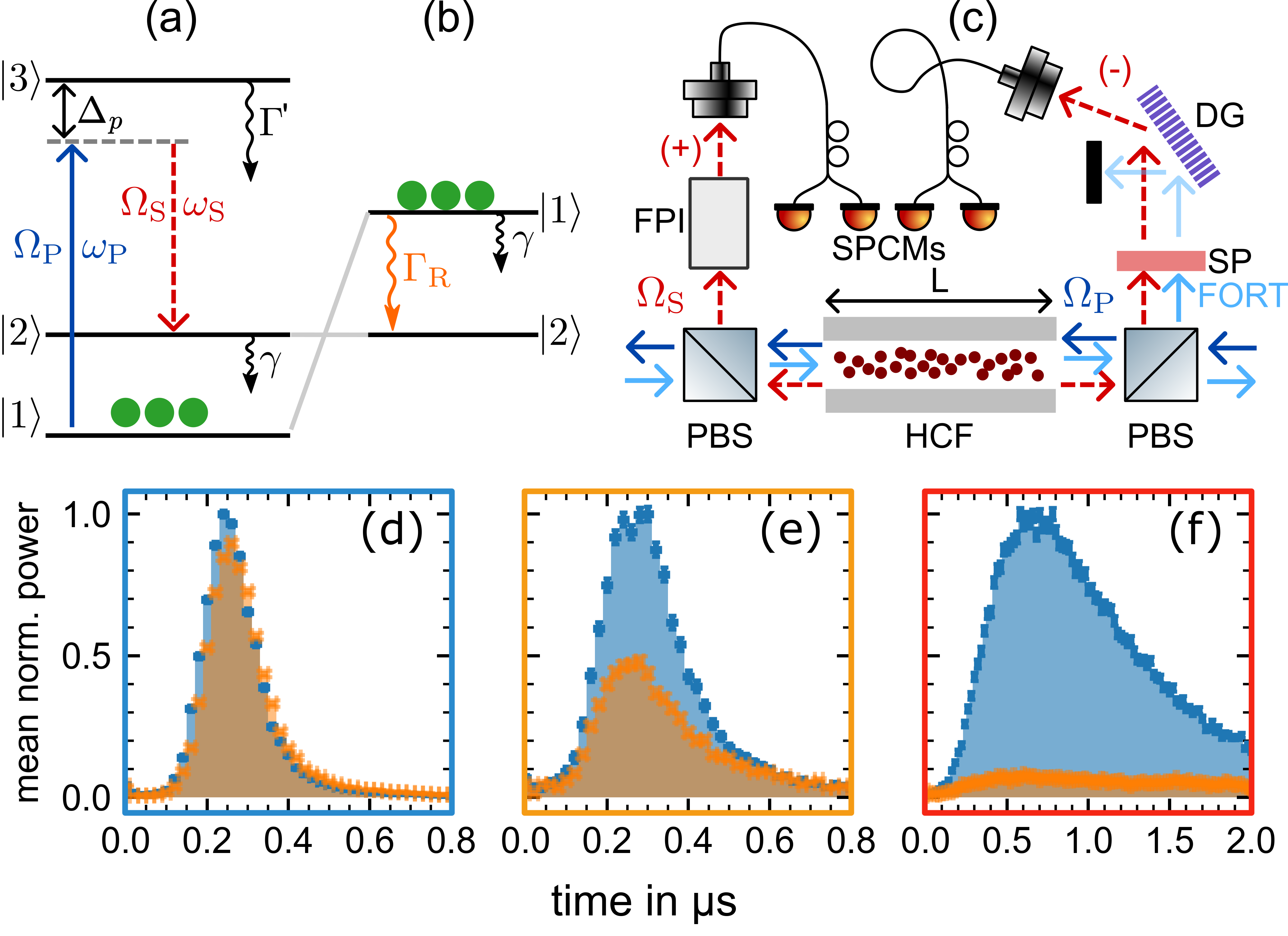}
\caption{
	\label{fig:eff_tls}
	Simplified three level system of $^{87}$Rb (a) with $|1\rangle,|2\rangle,|3\rangle\, \widehat{=}\, \mathrm{^2S_{1/2}\,F=1}, \, \mathrm{^2S_{1/2}\,F=2},\, \mathrm{^2P_{1/2}\,F'=2}$ that corresponds to an effective two-level system (b) for pump detuning $\Delta_p\gg\Gamma'$ of decay rate $\Gamma = \Gamma_R \propto \Omega_p^2$ with pump Rabi frequency $\Omega_p$.
    (c) Experimental setup. (d,e,f) Experimental examples for superfluorescent bursts measured in $(+)$ (blue) and $(-)$ direction (orange) for $(\sigma_\textrm{v} ; N_\textrm{mc})$ of $(1.5;160)$ (d), $(1.5;106)$ (e), and $(5;132)$ (f).}
\end{figure}

We here report the first experimental observation of collective emission from a disordered ensemble with controllable directionality in a \textit{non-chiral} one-dimensional (1D) waveguide.
Contrary to \cite{CMZ23}, there is no random spontaneous symmetry breaking.
In Figs.~\ref{fig:eff_tls}(d-f) we show three examples for the detected normalized light power emitted in forward $(+)$ and backward $(-)$ direction for different maximum cooperation number $N_\textrm{mc}$ and velocity spread $\sigma_\textrm{v}$ (see below for definitions). We clearly notice a strong asymmetry depending on the ratio of collective decay to motional timescale (see below) indicated by the box-color of the plots, although there is no chiral coupling or special spatial arrangement of individual emitters.
As we will show through comparison to numerical simulations, the directionality, only emerging in the collective dynamics, is caused by an interplay between atomic motion and the fact that the quantum emitters are effective two-level systems created in a Raman process [see Fig.~\ref{fig:eff_tls}(a)]. While the preferred direction is determined by the excitation geometry, the directionality itself is induced by atomic motion and requires a finite velocity spread. We note that directional Raman amplification in waveguide-coupled atomic ensembles has been discussed in \cite{LD19, PLJ22}. In these works, however, the directionality relies on chiral light-matter coupling.

Figure~\ref{fig:eff_tls}(c) depicts a simplified version of our experimental setup (for more details see Refs.~\cite{PYH21,SBP24}). 
We load up to $N = 2 \times 10^5$ laser-cooled $^{87}$Rb atoms from a magneto-optical trap into a hollow-core fiber [HCF, NKT Photonics, HC-800-02, core diameter  \SI{7.1}{\micro\meter}, $\textrm{NA}=0.092(6)$] where the atoms are guided by an optical far-off-resonant trap (FORT) operated in a running-wave configuration \cite{PYH21}. Inside the HCF the atomic ensemble has a temperature $T=1.1(1)$~mK and a near Gaussian radial density profile of $1/e$ radius $\sigma_a\sim 1.7~$µm, similar to the $1/e^2$ intensity mode radius $\sigma_p \sim 2.75$~µm of the laser beams. 
To avoid perturbing AC Stark shifts by the FORT during the measurements we switch it periodically off and on to provide measurement windows of $\tau_m = \SI{2.5}{\micro\second}$ duration during which the atomic ensemble is freely expanding [see Appendix (App.) \ref{sec:measurement_sequence}]. 
Before each measurement we prepare the ensemble in lower ground state $|1\rangle$  by optical pumping. 
We then send a pump pulse of power P, detuned by $\Delta_p = 26.4\Gamma'$ from transition $|1\rangle \leftrightarrow |3\rangle$, into the HCF which induces spontaneous Raman scattering to state $|2\rangle$ at a rate $\Gamma_R \propto \mathrm{P}/\Delta_p^2$. This coupling via Raman scattering creates an effective inverted two-level system $\{|1\rangle, |2\rangle\}$ \cite{SBP24} [see Fig.~\ref{fig:eff_tls}(a,b)] with a dipole moment that has a spatially oscillating phase factor $\sim e^{i k_p z}$, and whose single-atom decay rate $\Gamma = \Gamma_R$ can be tuned by adjusting $\mathrm{P}$. Here, the inversion arises naturally from the Raman process itself.

Due to the long-range coupling provided by the HCF, collective scattering can build up at the Stokes transition between states $|2\rangle \rightarrow |3\rangle$, leading to the emission of superfluorescent (SF) bursts. Since not all $N$ atoms contribute to collective scattering, we determine a maximum cooperation number $N_\mathrm{mc}^{(\pm)}= \eta_\mathrm{s} \eta_\mathrm{inh} \mu N$ taking into account the limited coupling of spontaneous radiation into each of the $(\pm)$ HCF modes ($\mu=\mathrm{NA}^2/4$), as well as spatial ($\eta_\mathrm{s}$) and spectral inhomogeneities ($\eta_\mathrm{inh}$) due to pump attenuation and AC Stark shifts by the pump \cite{SBP24}. This yields $N_\mathrm{mc}^{(\pm)}\lesssim 150$ effective collective emitters for each direction, resulting in a total $N_\mathrm{mc} = 2N_\mathrm{mc}^{(\pm)}$ (see App.~\ref{sec:MCN}). 
By tuning $\Gamma$ via the pump power $P_\mathrm{p}$, we control the relative (thermal) velocity distribution width $\sigma_\textrm{v} = \overline{v}/(\lambda_0\Gamma)$ of the ensemble, where $\bar{v}$ is the rms thermal velocity of the atoms and $\lambda_0$ the transition wavelength of the effective two-level system.
The ratio of collective emission to motional dephasing timescale is given by $R_\tau$ $= \tau_{th}/\tau_{col} $ $ =2\sigma_\textrm{v}/N_\textrm{mc}$.
Here, $\tau_{th} = \lambda / \bar{v}$ is the thermal dephasing time and $\tau_{col} =2/N\Gamma$ the characteristic superadiant time, neglecting directionality.
For varying  $\sigma_\textrm{v}$ and $N_\textrm{mc}$ we measure the light exiting the HCF in the $(\pm)$ directions, detected by a set of four single-photon counting modules (SPCMs, Excelitas, SPCM-AQRH-13) in a double Hanbury-Brown Twiss (HBT) configuration, followed by a time-tagging device (Swabian Instruments, TimeTagger 20). 
Details on the measurement protocols used to extract the relative $(\pm)$ amplitudes of the emitted bursts are given in App.~\ref{sec:measurement_strategies}. 

\begin{figure}[tb]
\includegraphics[width=1\columnwidth]{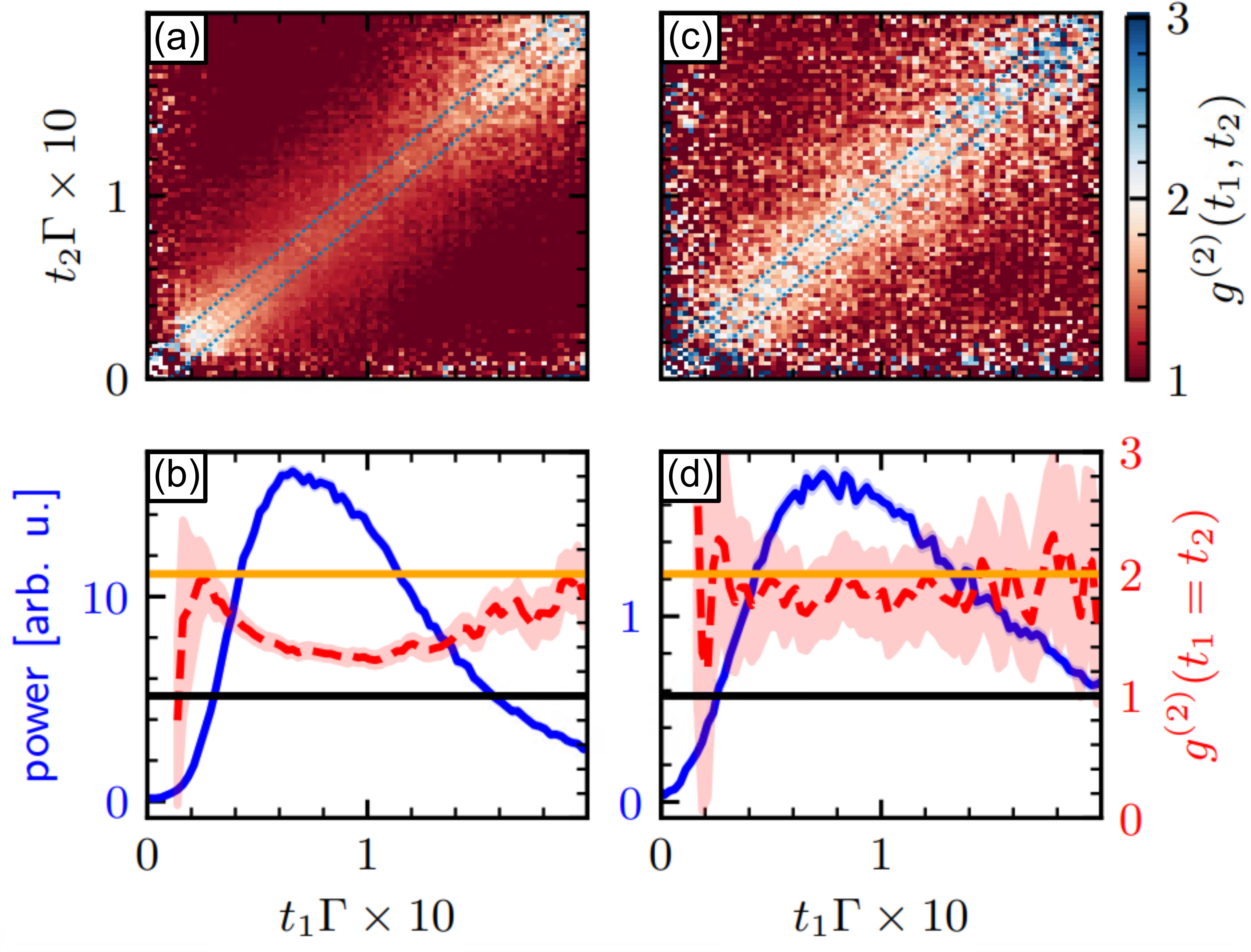}
\caption{
	\label{fig:correlations2D}
    Temporal evolution of the pulse power and correlation functions above (a,b) and close to the SF threshold (c,d) for N$_\textrm{mc} = 248(7)$ and N$_\textrm{mc} = 55(6)$, respectively.
	Two-time auto-correlation functions $g^{(2)}(t_1,t_2)$ are shown in (a,c) with the corresponding average count rates (blue) during the bursts along with the equal-time $g^{(2)}(t,t)$ (red, dashed). Intervals used for averaging $g^{(2)}(t,t)$ are marked by blue dotted lines in (a,c). Measured at $\Gamma = 2\pi\times 18.1(6)~$kHz, averaged over $20\times10^3$(a,b) and $50\times10^3$(c,d) realizations.
	}
\end{figure}

%
%
We first investigate the general pulse dynamics and correlations of the emitted bursts.
Figure~\ref{fig:correlations2D} shows the intensity profiles and the two-time correlation function
$g^{(2)}(t_1,t_2)$ for light emitted into the $(+)$ direction well above (a,b) and close to the threshold to SF emission (c,d).
We calculate the second-order correlation function $g^{(2)}(t_1,t_2) = \langle N_1(t_1) N_2(t_2) \rangle/(\langle N_1(t_1) \rangle\langle N_2(t_2) \rangle)$ from the measured data using the count rates $N_i$ of the $i$th SPCM. We apply averaging over different single experimental realizations.
We see in (a,b), the correlations are thermal with $g^{(2)}(t,t) \approx 2$ during the initial stages of the SF bursts, where dynamics are determined by vacuum fluctuations. As the bursts build up, $g^{(2)}(t,t)$ decreases, indicating a build up of coherence due to the synchronization of dipoles as observed and discussed in \cite{FFB25,BTL26}. The minimum value of $g^{(2)}(t,t)=1.31(5)$ occurs shortly after the burst reached its maximum. Thereafter, $g^{(2)}(t,t) \to 2$, indicating a loss of coherence. These observations are consistent with the Dicke model where the peak emission rate is reached when the population distribution between initial and final state is equal and the coherence is maximal \cite{D54, GH82}. The deviation from $g^{(2)}(t,t)$ reaching unity is expected for our extended sample and enhanced by shot-to-shot fluctuations of the burst delay inherent to the SF process.
Figure~\ref{fig:correlations2D}(c,d) shows similar data, but for conditions close to the SF threshold. In contrast to (a,b), we can see that $g^{(2)}(t,t) = 1.8(3)$, i.e., always close to the thermal value. As the emitted 
pulse exhibits a temporal width $1.5$ times larger than in (a,b) and obeys thermal statistics, we interpret this as amplified spontaneous emission \cite{MMS87}. 
We observe a smooth transition approaching the threshold, showing an increase in the minimal value of $g^{(2)}(t,t) \to 2$.
Measurements further below the threshold did not change the SF dynamics qualitatively but resulted in worse statistics.
Analogous measurements for the light emitted into the $(-)$ direction showed the same behavior. 
We note that the cross-correlation between the $(+/-)$ directions is close to unity except during the burst build-up (see Fig.~\ref{fig:CCFs} in App.~\ref{sec:app_CCF}.)

To understand the experimental results we employ numerical simulations based on the Truncated Wigner Approximation (TWA) for a spatially extended ensemble of spins \cite{mink2023collective}.
Collective light emission from an ensemble of atoms can be described by an effective spin model \cite{dung2002resonant,dzsotjan2011dipole,caneva2015quantum,asenjo2017atom}, obtained by integrating over the reservoir degrees of freedom of the light field. Here, the density matrix $\varrho$ of the spins obeys a Lindblad master equation, $\dot \varrho = - i [H,\varrho] + {\cal L}\varrho $, with effective Hamiltonian $H$ and Lindblad operator ${\cal L}$
\begin{align}
    H =& \frac{1}{2}\sum_{l,m} J_{lm}\, \sigma_l^+\sigma_m^-,\label{eq:H}\\
    {\cal L}\varrho =& \frac{1}{2}\sum_{l,m} \Gamma_{lm} \bigl(\sigma_l^+ \sigma_m^-\varrho + 
    \varrho \sigma_l^+ \sigma_m^- - 2 \sigma_m^-\varrho \sigma_l^+\bigr),\label{eq:dissipation}
\end{align}
where $\Gamma_{lm} \sim \textrm{Re}\bigl[{\vec d}_l^* \cdot \mathbf{G}(z_l,z_m,\omega_0)\cdot \vec d_m\bigr]$ describes the strength of the collective decay process and $J_{lm} \sim 
\textrm{Im}\bigl[{\vec d}_l^* \cdot \mathbf{G}(z_l,z_m,\omega_0)\cdot \vec d_m\bigr]$ the dipole-dipole exchange interaction.
$\vec d_m$ is the dipole matrix element of the atom at position $z_m$ and $\mathbf{G}(z_l,z_m,\omega_0)$ is the Green's tensor of the radiation field at  frequency $\omega_0$ of the atomic transition \cite{dung2002resonant}. 
In a 1D system, the collective coupling reads 
$\Gamma_{lm} = \Gamma_\textrm{1D} \bigl(\beta_+ e^{ik_0 z_{lm}} + \beta_- e^{-i k_0 z_{lm}}\bigr)$ and $J_{lm} = i \Gamma_\textrm{1D}/2  \bigl(\beta_+e^{i k_0 \vert z_{lm}\vert } - \beta_- e^{-i k_0 \vert z_{lm}\vert}\bigr)$, where the
 coefficients $\beta_\pm$ account for a potential chiral coupling of a single emitter to the waveguide, see e.g. \cite{pichler2015quantum,tebbenjohanns2024predicting,LTB24}. If 
$\beta_+=\beta_-=\frac{1}{2}$, there is full inversion symmetry between forward and backward direction.
$\Gamma_\textrm{1D}$ is the single atom emission rate into the waveguide, $k_0=\omega_0/c$ and $z_{lm} = z_l-z_m$ is the relative position of the atoms, which is time-dependent due to atomic motion. 
Additionally, we add single-atom decay with rate $\Gamma$ in Eq.~(\ref{eq:dissipation}).
The complex amplitude of the light field emitted in $(\pm)$ direction of the waveguide at the two ends is then given as a sum over emission contributions of all spins
$E_\pm = i\sqrt{\frac{\Gamma_\textrm{1D}}{2}} \sum_j e^{i(k_p\mp k_0) z_j} \, \sigma_j^-$, with $k_p=\omega_p/c$.

\begin{figure}[t]
\includegraphics[width=\columnwidth]{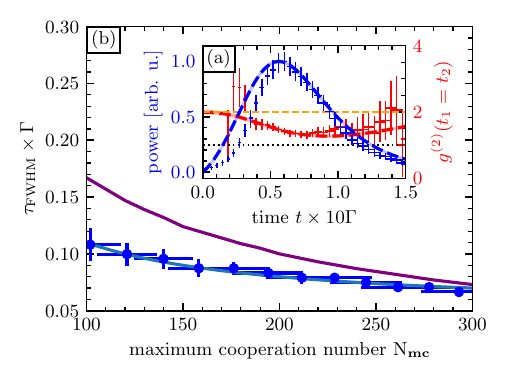}
\vspace*{-7mm}
\caption{
	\label{fig:correlations}
    (a) Emitted optical power vs time (blue) and equal-time $g^{(2)}(t,t)$ (red) vs $N_\mathrm{mc}$ in (+) direction for $N=71(6)\times10^{3}$ (corresponding to $N_\textrm{mc}=280$) and $\Gamma = 2\pi\times 33(2)~$kHz, averaged over $56\times 10^3$ realizations.
    Dashed lines show TWA simulations, while data points show measurement results. Horizontal lines indicate thermal (orange) and coherent (black) correlations.
    (b) Experimental SF burst widths (blue symbols) vs $N_\textrm{mc}$ for $\Gamma = 2\pi\times 33(2)~$kHz. 
    The lines show a least-squares fit of type $\tau_\mathrm{FWHM} =6.0(2)/(N_{\textrm{mc}}\Gamma)$ to the experimental data (light blue) and the results of full numerical calculations averaged over 5$\times 10^4$ trajectories (purple).
	}
\end{figure}

We numerically simulate the effective spin model, Eqs.(\ref{eq:H},\ref{eq:dissipation}),  using the 
TWA for spins \cite{MinkPRR2022,mink2023collective} and the collective correspondence rules \cite{mink2023collective}, which have proven to give accurate predictions of superradiance in waveguide systems \cite{tebbenjohanns2024predicting}.
The stochastic method is numerically 
inexpensive and allows for the simulation of up to several thousand interacting spins.
Details are given in App.~\ref{sec:AppA}.
As the total number of atoms $N\sim 10^5$ in the experiment is, however, larger, a straight-forward simulation is computationally too costly. We thus theoretically investigate a reduced system with a much smaller number of emitters 
$N_\mathrm{mc}\sim 300$ as in the experiment.
We compensate this by assuming stronger coupling $\Gamma_\mathrm{1D}^\prime =\Gamma$ to the waveguide mode, while keeping the cooperativity $N_\textrm{mc} \Gamma_{1D}^\prime = N \Gamma_\mathrm{1D}$ the same.
This clearly underestimates the effect of spontaneous emission events into non-waveguide modes in the simulation and other single-atom effects as dephasing, and thus does not allow to accurately describe the behavior near or below the threshold of SF emission. Above this threshold, the simulations of the reduced model do, however, reproduce the experimental results rather well for the temporal evolution of the burst power (blue) as well as the correlations (red) shown in Fig.~\ref{fig:correlations}(a). We observe qualitative accordance for all parameters in the collective decay.
Only the timescale, measured by the pulse widths shown in Fig.~\ref{fig:correlations}(b), is slightly larger than the experimental values.
However, both experiment and simulation exhibit similar (anti-proportional) scaling between $N_\textrm{mc}$ and $\tau_\mathrm{FWHM}$.

\begin{figure}[tb]
\includegraphics[width=1\columnwidth]{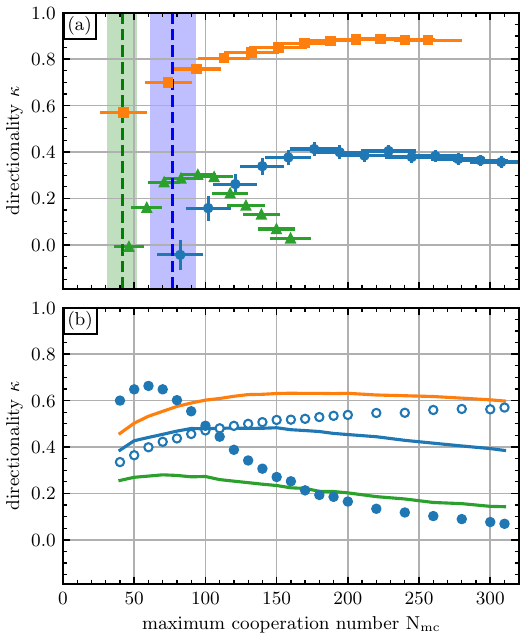}
\caption{
	\label{fig:fwd_bwd_data}
	(a) Measured directionality $\kappa$ (symbols) of the SF bursts amplitudes in $(\pm)$ direction vs $N_\mathrm{mc}$. The pump power was set to three different values at $\Delta_p = 26.4\Gamma_{D_1}$ corresponding to $\Gamma_R / (2\pi) = 20(2)~$kHz, $33(2)$~kHz, $67(2)~$kHz. This results in $\sigma_\mathrm{v}$ of 5.0(2)(\textcolor{orange}{\scriptsize$\blacksquare$}), 3.0(1)(\textcolor{RoyalBlue}{$\bullet$}) and 1.50(2)(\textcolor{OliveGreen}{$\blacktriangle$}) respectively. The vertical lines indicate the SF thresholds, respectively.
    (b) Corresponding $\kappa$ as obtained from the TWA simulations including motion (lines). The results of the static effective model for $\sigma_\mathrm{v} = 3$ are shown by symbols,
    assuming constant $\tau$ (\textcolor{RoyalBlue}{$\circ$}) as well as $\tau(N_\textrm{mc}) = (N_\textrm{mc}\Gamma/2)^{-1}$ (\textcolor{RoyalBlue}{$\bullet$}). Each data point was averaged over $5\times 10^4$ trajectories, resulting in uncertainties smaller than the symbol size.
    }
\end{figure}

We now turn to a detailed investigation of the directionality of the SF emission shown in Fig.~\ref{fig:eff_tls}(d-e). In order to measure the directionality more precisely we change the direction of the pump beam instead of using differing detection ports (see App.~\ref{sec:measurement_strategies}).
We quantify the directionality by $\kappa = (R_+-R_-)/(R_++R_-)$, where $R_\pm$ are the peak photon rates in $(\pm)$ direction.
The experimental dependence of $\kappa$ on $N_\textrm{mc}$ is shown in Fig.~\ref{fig:fwd_bwd_data}(a) for three different values of the velocity spread $\sigma_\mathrm{v}$ (in units of $\lambda_0 \Gamma$).
In each measurement we observe the following behavior: At low $N_\textrm{mc}$, the emission loses directionality, i.e., $\kappa \rightarrow 0$. Vanishing directionality can be seen for the two lowest $\sigma_\mathrm{v}$ (green, blue). For the largest $\sigma_\mathrm{v}$ (orange) we were not able to reach sufficiently small $N_\mathrm{mc}$. With increasing $N_\textrm{mc}$, $\kappa$ first increases towards a maximum, before decreasing again for  further increasing $N_\textrm{mc}$. (This is also the case for the orange data points as closer inspection reveals.) For the maximal $N_\textrm{mc}$ available in our experiment, $\kappa$ approaches zero only for the smallest $\sigma_\mathrm{v}$ (green circles). A decrease for large $N_\textrm{mc}$ can be seen for all measurements, however. 
In Fig.~\ref{fig:fwd_bwd_data}(b) we show the results of the simulations.
Most importantly, we took atomic motion during the emission process into account. We assumed a Maxwell-Boltzmann distribution of velocities with width $\sigma_\mathrm{v}$, as determined from experimental data.
Without atomic motion, the simulations yield $\kappa =0$.
As can be seen in Fig.~\ref{fig:fwd_bwd_data}(b), including atomic motion indeed results in a non-uniform collective emission in $(+)$ and $(-)$ direction with $\kappa > 0$. 
The simulations reproduce the same behavior as described for the experimental data. 
However, we only observe good quantitative agreement for $\sigma_\textrm{v}=3$ (blue).
This can be explained as follows.
For small $N_\textrm{mc}$, the increase in $\kappa$ reflects the interplay between purely spontaneous emission and the emergence of collective emission beyond the threshold.
As argued before, the model used in the simulation with smaller atom number $N_\textrm{mc}$ yields only a faithful representation of collective effects but strongly underestimates spontaneous emission and amplified spontaneous emission below threshold.
Hence, disagreement at low $N_\textrm{mc}$ is to be expected. Furthermore, the maximum directionality is
amplified by the difference in SF thresholds, an effect not correctly represented in the full simulations.
This becomes more relevant when the threshold difference is large, i.e., at large $\sigma_\textrm{v}$. 
Finally, the simulations neglect the Stokes gain of the ensemble \cite{SBP24}, which also contributes to the directionality. This effect can account for deviations of up to $7\%$ (see App.~\ref{sec:Stokes-gain}).
Given the strong simplifications of the model, a precise quantitative agreement cannot be expected.
However, the qualitative dependencies of $\kappa$ on both, maximum cooperation number $N_\textrm{mc}$ and velocity spread $\sigma_\textrm{v}$ are well described by the model.

We now argue that the directionality and its dependence on $N_\textrm{mc}$ and $\sigma_\textrm{v}$ is due to motional induced dephasing and the fact that the dipole transition matrix elements of the dressed two-level system [see Fig.~\ref{fig:eff_tls}(a)] have a space-dependent phase. 
To this end, we describe the effect of atomic motion by a static model
with position uncertainty $z_l \to z_l +\xi_l$, where the $\xi_l$ are random with $\overline{\xi_l}=0$, $\overline{\xi_l^2} = \sigma_z^2 $.
Since the position uncertainty results from the motion of the atoms during the characteristic time $\tau$ of the collective coupling process one has $\sigma_z = \overline{v}\, \tau$ \footnote{Note that different from \cite{kusmierek2024emergence} which discusses a related mechanism for emergent unidirectionality, the spatial separation between all emitters has the same uncertainty $\sigma_z$ and does not accumulate.}.
The static model strictly applies only, when the spatial phase factor of the effective dipole $\sim e^{i k_p z}$ 
changes substantially during the time $\tau$. It nevertheless serves as an instructive toy model to explain the physical origin
of the induced directionality of the collective emission.
Averaging $\Gamma_{lm}$, $J_{lm}$, and $E_\pm$ over the spatial distribution creates an asymmetry in the coupling in $(\pm)$ direction, where $\beta_+ = \frac{1}{2}$ and $\beta_- = \frac{1}{2} \exp\bigl\{- \bigl(4 \pi \frac{\overline{v} \tau}{\lambda_0}\bigr)^2\bigr\}$.
%
%
%
%
Above, we have used that $k_p\approx k_0=2\pi/\lambda_0$.
Furthermore, the averaged emitted light intensity in $(-)$ direction $I_-\sim \sum_{jl} \overline{e^{i(k_p+k_0)z_{jl}}} \langle \sigma^+_j \sigma^-_l\rangle$ attains a term $\beta_-^2$ in the contributions with $j\ne l$  
while $I_+$ remains approximately unchanged.
Thus, for sufficiently large $\sigma_z=\bar{v}\tau$, collective emission into the $(-)$ direction is suppressed.
This way, the static model maps the velocity distribution onto a location blur of width $\sigma_z$. 

In Fig~\ref{fig:fwd_bwd_data}(b) we also show two examples for TWA simulations of a static ensemble of atoms with effective couplings $\beta_\pm$ and $\sigma_\textrm{v}=3$. 
In the first example, a constant $\tau = 0.019\,\Gamma^{-1}$ (open circles) is chosen and in the second we use
the characteristic superradiant time \cite{GH82} $\tau(N_\textrm{mc}) = (N_\textrm{mc}\Gamma/2)^{-1}$ (full circles) (see App~\ref{sec:static_model}). 
A constant $\tau$ neglects the fact that for increasing $N_\textrm{mc}$ the collective timescale shortens. 
On the other hand $\tau(N_\textrm{mc})$ overestimates the impact of the varying collective timescale especially for small $N_\textrm{mc}$.
In lack of a precise theoretical prediction for $\tau$ one 
could use a fit of the time scale for each $N_\textrm{mc}$, denoted as $\tau^*(N_\textrm{mc})$, such that the static and the dynamic model agree.
The matched $\tau^*(N_\textrm{mc})$ and a comparison to $\tau(N_\textrm{mc})$ can be found in Fig.~\ref{fig:tau_factors} of the App.~\ref{sec:static_model}. 
Our interpretation of the experimental findings is thus as follows:
At low $N_\textrm{mc}$, beneath the threshold to collective decay (see App.~\ref{sec:MCN}), indicated by vertical lines in Fig.~\ref{fig:fwd_bwd_data}(a), the emission results from incoherent, (amplified) spontaneous decay and therefore exhibits no directionality. Note that at the lowest $P_\mathrm{p}$ ($\sigma_\mathrm{v}=5$) no clear threshold towards the SF regime could be determined because the transition is less sharp. 
Once the threshold is crossed, collective effects become significant and directionality in $(+)$ direction arises.
In the collective regime, the ratio of collective to motional timescale $R_\tau$ provides a qualitative guide to the behavior of $\kappa$.
If $R_\tau$ increases, the collective processes slow down compared to the thermal motion and the resulting dephasing results in stronger directionality. This can be achieved by decreasing $N_\textrm{mc}$ or increasing $\sigma_\textrm{v}$.
Vice versa, increasing $N_\textrm{mc}$ results in decreasing directionality deep in the collective regime. Depending on $\sigma_\mathrm{v}$ more or less collective emitters are required to counter-act the thermal dephasing. Only for the smallest $\sigma_\mathrm{v} = 1.5$ we reach the regime where collective dynamics outpace motional dephasing and the emission becomes symmetrical again. 

Additionally, in our system the maximal directionality near the SF threshold is increased by the fact that the threshold in $(-)$ direction is at slightly higher $N_\textrm{mc}$ than in the $(+)$ direction due to the additional decoherence introduced by two-photon Doppler broadening, estimated as $\approx 0.2\,\Gamma'$.
The latter yields a larger threshold difference, i.e., a stronger increase of $\kappa$, increasing the maximum directionality up to 
$\kappa = 0.89(1)$.
We can thus easily tune the collective emission directionality in our system by the pump power and atom number, i.e., the collective decay rate.

In summary, we have given clear experimental evidence of an emergent directionality of SF collective emission in a \textit{non-chiral} waveguide, tunable by the ratio of collective emission rate to the rate of motional dephasing. 
The experimental observations are in fair qualitative agreement with full-scale numerical TWA simulations if the motion of atoms during the emission process is taken into account. A simple qualitative understanding can be obtained by replacing the model including atomic motion by a model for atoms at rest with a localization uncertainty. 
For the directionality to occur, a breaking of the $(+/-)$ symmetry is needed. This is provided here by the pump laser parallel to the fiber axis. If the elongated ensemble could be excited from a different angle--not perpendicular to the HCF--motion-induced directional emission would also occur. Similar directional collective emission could be observed without a pump laser in a thermal ensemble of initially inverted two-level emitters having a dipole moment with a spatially oscillating phase. Our work identified a new mechanism of emerging directionality of collective radiative couplings in a system with fully isotropic emitters. 
Besides being of fundamental interest on its own, we anticipate various potential applications in photonics \cite{HWA13,WZG13}, waveguide QED \cite{SPI23}, and new ways to realize directional  interactions in quantum optical systems \cite{fruchart2021non,CAT18,KKR21,NCK22,NBB25} using isotropic building blocks and collective phase engineering.


\begin{acknowledgments}
This work was funded by the Deutsche Forschungsgemeinschaft (DFG, German
Research Foundation) – Project No. 410249930 and SFB/TR 185, Project No.277625399.
\end{acknowledgments}

YS and BS conducted and evaluated the experiment, JH performed the TWA simulations, the static model was developed by MF. TP supervised the experiment with support of TH and MF the theoretical work.
All authors contributed to the discussion of the results and the writing of the manuscript.

\section*{Data Availability Statement}
The data used for this work are available from the authors upon reasonable request. 

\nocite{MinkPRR2022,mink2023collective}
\nocite{MinkPRR2022}
\nocite{mink2023collective}
\nocite{PhysRevResearch.4.023026}
\nocite{GH82}
\nocite{SBP24}
\nocite{PYH21}
\nocite{philip_thomas_starkey_software_2019}
\nocite{PSV79}


\appendix
\renewcommand{\thefigure}{A\arabic{figure}}
\setcounter{figure}{0}

\renewcommand{\thetable}{A\arabic{table}}
\setcounter{table}{0}

\section{TWA simulations}
\label{sec:AppA}

\subsection{General approach}

The Truncated Wigner Approximation (TWA) for spins is a semiclassical method to calculate the time evolution of the density matrix $\rho$ of a large ensemble of spins \cite{MinkPRR2022,mink2023collective}. The idea is to shift the description 
from Hilbert space into phase space and truncate the resulting equation of motion 
of the phase space representation of $\rho$, the Wigner function,
to a Fokker-Planck equation (FPE). These equations have equivalent stochastic differential equations (SDEs) which allow efficient numerical simulations while at the same time the system size can be extended to thousands of spins.
All operators $\hat A$, including the density operator, are expressed in terms of products of phase-point operators $\hat \Delta$ 
\begin{eqnarray}
 \hat A = \int\! d\Omega\, {\cal A}(\Omega)\,  \hat{{\Delta}}({\Omega}),
\end{eqnarray}
with the corresponding Weyl symbol $\mathcal{A}(\Omega)$
and the integration over two angles $\theta$ and $\phi$ for every spin, representing it in phase space. 
\begin{eqnarray}
    \hat{\Delta}(\theta, \phi) =
    \begin{pmatrix}
        1 + \sqrt{3} \cos \theta & \sqrt{3} e^{-i \phi} \sin \theta \\
        \sqrt{3} e^{i \phi} \sin \theta & 1 - \sqrt{3} \cos \theta
    \end{pmatrix}.
\end{eqnarray}
The Lindblad equation for $\rho$ is translated into phase space by making use of the fact that
any product of spin operators with $\rho$ can be expressed in terms of the basis
\begin{eqnarray}
    \hat \Delta(\Omega),\enspace \partial_\theta  \hat \Delta(\Omega), \enspace  \partial_\phi  \hat \Delta(\Omega), \enspace  \partial^2_\phi  \hat \Delta(\Omega),
\end{eqnarray}
which gives rise to correspondence rules between Hilbert space and phase space.
These lead to a partial differential equation for the Wigner function \cite{MinkPRR2022}, which upon truncation of higher-order terms can be approximated by a FPE. 
To describe an entirely collective system we use the collective correspondence rules \cite{mink2023collective} which include the appropriate approximation of the equations of motion already. 

The Lindblad master equation for $N$ collectively-coupled spins~\eqref{eq:H} and \eqref{eq:dissipation} in the main text, yields the following SDEs
%
%
%
%
%
\begin{align}
    \mathrm{d} \theta_n =& \, \bigg\{ \sqrt{3} \, \sum_{m=1}^N \sin{\theta_m} \bigg[ \sin{\phi_{mn}} \left( J_{mn}' + \frac{\Gamma''_{mn}}{2} \right) \nonumber \\
    &+ \cos{\phi_{mn}} \left( \frac{\Gamma'_{mn}}{2} + J_{mn}'' \right) \bigg] + \frac{\Gamma_{nn}}{2} \cot{\theta_n} \bigg\} \, \mathrm{d} t \nonumber \\
    &+ \sum_{m=1}^N \bigg[ \left( - G_{nm}' \cos{\phi_n} + G_{nm}'' \sin{\phi_n} \right) \, \mathrm{d} W_{\theta_m} \nonumber \\
    &\quad + \left( G_{nm}' \sin{\phi_n} + G_{nm}'' \cos{\phi_n} \right) \, \mathrm{d} W_{\phi_m} \bigg] ,  \\[1em]
    \mathrm{d} \phi_n = & \, \bigg\{ \sqrt{3} \sum_{m=1}^N \sin{\theta_m} \bigg[ \cos{\phi_{mn}} \left( -J_{mn}' + \frac{\Gamma''_{mn}}{2} \right) \nonumber \\
    &+ \sin{\phi_{mn}} \left( \frac{\Gamma_{mn}'}{2} - J_{mn}'' \right) \bigg] \bigg\} \cot{\theta_n} \, \mathrm{d} t \nonumber \\
    &+ \sum_{m=1}^N \cot{\theta_n} \bigg[ \left( G_{nm}' \sin{\phi_n} + G_{nm}'' \cos{\phi_n} \right) \, \mathrm{d} W_{\theta_m} \nonumber \\
    &\quad + \left( G_{nm}' \cos{\phi_n} - G_{nm}'' \sin{\phi_n} \right) \, \mathrm{d} W_{\phi_m} \bigg], 
\end{align}
where the coupling matrices
\begin{eqnarray}
    \Gamma &=& \Gamma' + i \, \Gamma'', \\
    \mathrm{J} &=& \mathrm{J}' + i \, \mathrm{J}''
\end{eqnarray}
are split up in their real and imaginary parts and $\Gamma = GG^T$. The terms $\mathrm{d} W_\mu$ are increments of independent Wiener processes.

The additional single particle terms can be treated separately. 
Here we only have spontaneous decay with rate $\Gamma$ which does not contribute to the collective mode. This gives the additional terms
%
%
%
%
\begin{align}
    \mathrm{d} \theta_n =& 2 \Gamma \Big( \cot{\theta_n} + \frac{1}{\sqrt{3}} \csc{\theta_n} \Big) \mathrm{d}t,\\
    \mathrm{d} \phi_n =& 2 \sqrt{\Gamma \big( 1 + 2 \cot^2{\theta_n} + \frac{2}{\sqrt{3}} \csc{\theta_n} \cot{\theta_n} \big)} \, \mathrm{d}W_n.
    \end{align}
In the waveguide the cooperative emission of the atoms is described by a collective coupling to the 1D continuum of modes. 
Thus the hermitian matrices for collective interaction $J_{mn}=J_{nm}^*$ and decay $\Gamma_{mn}=\Gamma_{nm}^*$ simplify to 
\begin{eqnarray}
    && J_{mn} - \frac{i}{2}\Gamma_{mn} = \quad g_{mn} \\ 
    &&= -i\frac{\Gamma_\textrm{1D}}{2}\exp{\Bigl( \frac{2\pi i}{\lambda_0} | z_n - z_m | \Bigr)} \cdot \exp{\Bigl(-\frac{2 \pi i}{\lambda_p} ( z_n - z_m )\Bigr)}\nonumber,
\end{eqnarray}
with $\Gamma_{1D}$ being the single atom coupling strength to the collective mode, $\lambda_0 = \frac{2 \pi}{k_0}$ the wavelength of the two-level transition and $\lambda_p = \frac{2 \pi}{k_p}$ the wavelength of the pump light. 

The emitted electric field is then given as the weighted sum over the single spins \cite{PhysRevResearch.4.023026} plus the phase contribution from the pump field:
\begin{align}
    \hat{E}_\pm(z_d, t) = i \sqrt{\frac{\Gamma_{1D}}{2}} \sum_{n=1}^N  \, \exp{\Bigl(i (k_p\mp k_0) z_n\Bigr)} \, \hat{\sigma}_n^-(t).
\end{align}

With the equations above we can now simulate the system with the following steps:
\begin{enumerate}
    \item Sample the  position of each spin uniformly along the $z$-axis.
    \item At $t=0$ the spins start in the excited state. Therefore, $\theta$ is fixed at $\theta_i = \arccos\left(\frac{1}{\sqrt{3}}\right)$ while $\phi_i$ is randomly sampled for each spin between $0$ and $2 \pi$.
    \item Calculate the coupling matrices and use the SDEs to determine the time evolution and Weyl symbols of the spin operators.
    \item Repeat steps $1-3$ for $\sim 10^5$ independent trajectories.
    \item Calculate the system's observables by averaging over the trajectories of the Weyl symbols.
\end{enumerate}
For every numerical computation we used the Julia programming language v1.11 with the \textit{DifferentialEquations} package because of its optimized performance and high stability for a large collection of SDE solvers while at the same time it is easily possible to handle large ensembles of trajectories. In our specific case we used the Euler-Maruyama algorithm with a fixed step size of $dt = 10^{-3}\Gamma^{-1}$.

To resemble the experiment we must add atomic motion. Therefore, in each timestep $\mathrm{d}t$ the position $z_n$ of each atom is changed by $z_n' = z_n + \nu_n \mathrm{d}t$. The $\nu_n$ are for each trajectory drawn from a Gaussian distribution with mean value zero and standard deviation $\sigma_\mathrm{v}$ which was defined in the main text. The computational effort increases because the coupling matrices now have to be calculated in each timestep of every trajectory.

%
\subsection{Static model}
\label{sec:static_model}

In the static model the atoms have a fixed position for each trajectory but a positional blur as explained in the main text. Averaging then leads to the following coupling matrices
\begin{align}
   \overline{\Gamma}_{lm} =& \quad \frac{\Gamma_\textrm{1D}}{2} \Bigl(e^{i k_0 z_{lm}} + \beta_- \,  e^{-i k_0 z_{lm}}\Bigr),\label{eq:Gamma_eff}\\
   \overline{J}_{lm} =& \quad \textrm{sgn}(z_{lm})\frac{\Gamma_\textrm{1D}}{4i} \Bigl(e^{ik_0 z_{lm}} - \beta_- \,  e^{-ik_0 z_{lm}}\Bigr),\label{eq:J_eff}
\end{align}
with the suppression factor
\begin{eqnarray}
    \beta_- = \exp{ \left\{ -\bigg( \frac{4 \pi \bar{v} \tau}{\lambda_0} \bigg)^2 \right\}}.
\end{eqnarray}
As before, $\bar{v}$ is the rms width of the velocity distribution, set to match the experimental one, and $\tau$ is the freely adjustable parameter mapping the velocity distribution onto a position blur of width $\tau\bar{v}$. It represents the time scale on which atomic motion-induced dephasing affects the collective coupling in the static model.

\begin{figure}[tb]
\includegraphics[width=1\columnwidth]{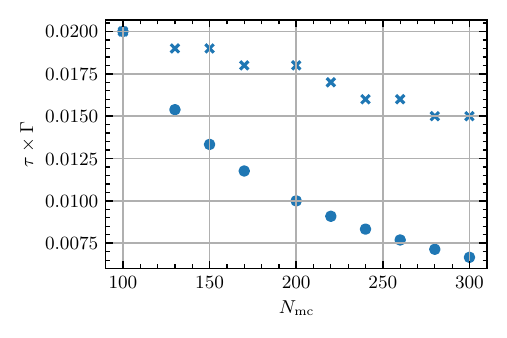}
\caption{
   Timescales $\tau^*(N_\textrm{mc})$ (crosses) and $\tau(N_\textrm{mc})$ (full circles) in units of the spontaneous emission rate $\Gamma$ vs $N_\mathrm{mc}$ used to map the dynamic onto the static model for $\sigma_\mathrm{v} = 3$. For $\tau^*$ we achieve the best match between static and dynamic model in Fig.~\ref{fig:fwd_bwd_data}(b) while $\tau(N_\textrm{mc}) = (N_\textrm{mc} \Gamma/2)^{-1}$ is the characteristic superradiant time of the system.
   }
\label{fig:tau_factors}
\end{figure}

In Figure \ref{fig:tau_factors} we show the timescale $\tau = \tau^*(N_\textrm{mc})$ required in the static model to reproduce the same directionality $\kappa$ at $\sigma_\mathrm{v} = 3$ as in Fig.~\ref{fig:fwd_bwd_data}(b) (blue line), together with the characteristic superradiance time $\tau(N_\mathrm{mc})$ defined in \cite{GH82}. Comparing Figs.~\ref{fig:fwd_bwd_data}(b) and \ref{fig:tau_factors} we see that for small cooperative atom numbers $(N_\textrm{mc})$ the system is not yet determined solely by collective emission rate but by the dephasing rate $\gamma$ and spontaneous emission rate $\Gamma$. On the other hand, for large atom numbers, the role of superradiance time appears to be overestimated. This can be understood by noting that the assumption that $N_\textrm{mc}/2$ atoms emit into each direction neglects the directional dependence of the emission.

\section{Experimental Methodology}
\label{sec:AppB}

\subsection{Measurement Sequence}
\label{sec:measurement_sequence}

We load the HCF every $1.25$~s from the magneto-optical trap (MOT) as described in \cite{PYH21}, aiming to maximize the number of atoms loaded into the HCF. 
Timing of the laser pulses and positioning of the MOT cloud (magnetic fields) are controlled by a 24-channel FPGA-based programmable pulse generator (SpinCore, PulseBlaster PB24-100-4k-PCIe) and a digital-to analog output card (National Instruments, PCI-6110) both addressed via homemade software based on the labscript-suite environment developed in \cite{philip_thomas_starkey_software_2019}.
\begin{figure}[tb]
\includegraphics[width=1\columnwidth]{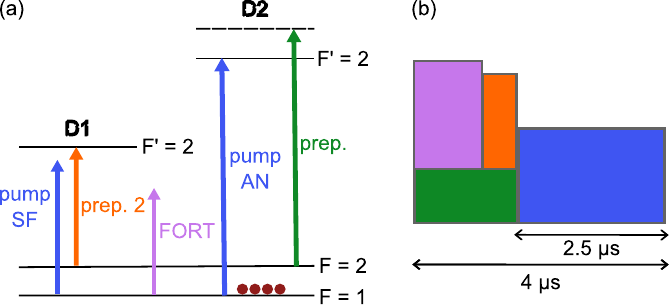}
\caption{
	\label{fig:Measrement_Prep_Scheme}
	(a) Coupling scheme used during preparation measurements. (b) Pulse sequence of a single measurement. We apply a detuned pump beam on the D$_1$ line for SF measurements, while we use a resonant pump beam on the D$_2$ line to determine the atom number.}
\end{figure}
As collective scattering requires the buildup of coherence, we must reduce decoherence on the time scale of the SF process. The major source of decoherence are inhomogeneous AC Stark shifts created by the strong FORT guiding the atoms inside the HCF. For a FORT depth of around $8$~mK the ground state frequency is shifted by around $14\Gamma_{D_1}$. This results in an ensemble temperature of $\approx 1~\textrm{mK}$ or 13\% of the potential depth, which is typical. This temperature was determined from
transverse time-of-flight measurements, verified in \cite{PYH21} to
agree in axial and radial directions. Both FORT intensity and atomic density distribution exhibit similar radial widths thus resulting in large inhomogeneities. We therefore periodically switch the FORT off and on at a frequency of $250~$kHz to provide measurement windows of duration $\tau_m = 2.5~$µs during which the atomic ensemble is freely expanding (while the atoms are trapped again during the remaining $1.5$~µs). This avoids AC Stark shifts by the FORT while inducing almost negligible transit-time broadening. 
These measurement windows are sufficiently short to allow for repeated interaction with the same atomic ensemble, while losses occur on a time scale of 1 ms only.  
We could use 650 measurement cycles before loading the HCF again from the MOT.
Before each measurement, the atomic ensemble has to be prepared in the initial state $|1\rangle$ by optical pumping, i.e., we must empty the metastable state $|2\rangle$. To this end, we first activate an off-resonant preparation beam close to transition $|F=2\rangle \rightarrow |F=2'\rangle$ of the D$_2$ line whilst for $2.4~$µs the FORT is still on. The AC Stark shift induced by the latter brings this transition close to resonance. As soon as the FORT is off, a second, resonant preparation beam on transition $|F=2\rangle \rightarrow |F=2'\rangle$ on the D$_1$ line is applied together with the first preparation beam for a duration of 300~ns 
(see Fig.~\ref{fig:Measrement_Prep_Scheme}). Both preparation beams are switched off before the off-resonant pump beam of the SF measurements is switched on.

\subsection{Measurement protocols}
\label{sec:measurement_strategies}
During our measurements we employed two measurement protocols A and B, where A mainly served to ensure the validity of B.
Protocol A aims at simultaneously detecting light emitted into opposite directions along the HCF axis for a pump beam of fixed direction and power. In particular, this requires detecting signals both co- and counter-propagating with respect to the strong FORT beam. The latter is approximately six orders of magnitude more intense than the signal of interest.
Since all pump and signal beams spatially overlap with the intense FORT inside the HCF, spectral separation is required before detection. This was done by propagating the light leaving the HCF in direction of the FORT onto a shortpass filter (Edmund Optics SP $\SI{800}{\nano\meter}$ 64333) followed by a grating (Thorlabs GR25-1208) before detecting the signal using two of the four SPCMs available. This setup permitted determination of relative delays of the bursts in both $(\pm)$ directions. The results show that the bursts in $(\pm)$ direction occur simultaneously. 
While providing precise information on the relative timing in $(\pm)$-direction, this protocol does not allow for determination of the relative burst intensities in opposite directions due to the uncertainties in the transmission.
\newline\noindent
We therefore employed protocol B where we altered the pump beam direction sequentially while using only a single detection channel. Depending on the direction of the pump beam, the light detected was then emitted into either the $(+)$ or $(-)$ direction. Here, we ensured equal pump power during subsequent measurements using the mean delay of the bursts. This delay scales with $\Gamma$ \cite{PSV79,SBP24}, i.e., the pump power. Combined with the knowledge obtained by protocol A, this method proved to be more reliable than characterizing all transmissions and coupling efficiencies, yet still resulting in large uncertainties. 
All results presented in the main text were obtained using protocol B. 

\subsection{Tunable decay rate}
One of the main differences to other experiments studying collective scattering is the ability to control the decay rate in our effective two-level system by varying the pump power and detuning.
Here, the single-atom decay rate is given by the spontaneous Raman scattering rate $\Gamma_R=1/2\Gamma' \Omega_p^2/(\Delta_p^2+2S)$, with the AC Stark shift $S$ of the transition frequency.
This decay rate can be much smaller than the excited state decay rate $\Gamma'$, therefore enabling the study of decay on much longer and more easily accessible time scales.
As discussed in \cite{SBP24}, we employ a radially-averaged effective decay rate $\langle \Gamma_R\rangle_r$ to account for the inhomogeneous pump intensity and atomic density distributions.\newline\noindent
The range of explorable decay rates is upper bound by the fact that at high pump power, $\Gamma_R$ does not grow monotonically because the AC Stark shift also increases with the pump power. As a result, one can estimate $\Gamma_{R,max}=\Gamma'/16$ . In principle, the lower bound is $\Gamma_{R,min}=0$, but experimentally we are limited by the increasing SF timescale with decreasing $\Gamma_R$ which eventually exceeds the duration of our measurement window, beyond which the freely expanding ensemble begins to collide with the fiber walls. 
In the measurements presented, the decay rate was varied over the range $\Gamma_R/(2\pi) = \langle \Gamma_R\rangle_r/(2\pi) \approx$ 18~kHz to 67~kHz. 

\subsection{Determination of the maximum cooperation number}\label{sec:MCN}

The maximal cooperation number $N_\textrm{mc}$, i.e., the effective number of collective emitters, is a key parameter to determine the collective decay rate. It depends on the total atom number $N$, as well as the pump power $P_p$ and the detuning $\Delta_p$ as shown in \cite{SBP24}.
Here it was shown that using $N_\textrm{mc}^{(\pm)}$ instead of $N$ recovers the scaling of collective decay as known for homogeneous ensembles, despite the present inhomogeneities. Notably, when detecting emission into a single direction, only the fractional solid angle $\mu$ in this direction is relevant and not the total fractional solid angle $2\mu$ of the $(\pm)$ modes of the waveguide. The latter leads to the conclusion that $N_\mathrm{mc} =2N_\textrm{mc}^{(\pm)}$ is required for correct modeling of bidirectional emission. 
This is equivalent to accounting for working with $\Gamma_\textrm{1D} = \Gamma$ instead of $\Gamma_\textrm{1D}^\pm = \mu\Gamma$. As the characteristic superradiant time $\tau$ is given by $\Gamma_\textrm{1D}^\pm N $ \cite{GH82}, this yields $\tau(N_\mathrm{mc}) = (N_\mathrm{mc}\Gamma/2)^{-1}$.
It is also noteworthy that since the inhomogeneities scale with pump power and detuning varying $N_\textrm{mc}$ can be obtained for the same number of atoms inside the HCF.

Since SF and atom number cannot be measured simultaneously, we experimentally characterize the total number of atoms $N$ loaded into the HCF by time-resolved optical pumping \cite{PYH21} before each SF measurement. To this end, a resonant pump beam on the D$_2$ line is employed instead of the off-resonant pump used for SF (see Fig.~\ref{fig:Measrement_Prep_Scheme}). Measuring the transmitted power of the resonant pump with and without atoms in the HCF yields the power absorbed by the atomic ensemble and therefore $N$. 

\subsection{Determination of the SF threshold}
\label{sec:threshold}
\begin{figure}[t]
\includegraphics[width=\columnwidth]{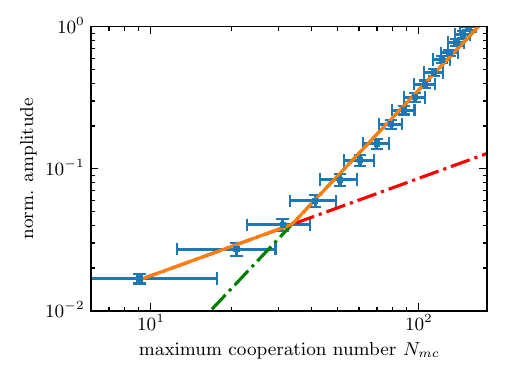}
\caption{
	\label{fig:example_threshold}
	SF burst amplitude vs $N_\mathrm{mc}$ used to determine the threshold to collective decay. The maximum of two linear slopes
    was used as fit function (orange), with one representing the
    non-collective (red) and the second one the collective scaling
    (green). The slope in the SF regime is 2 and in the non-
    collective regime 0.69 for the data shown.}
\end{figure}

Collective decay only occurs if the atomic dipoles can synchronize with each other. This requires the collective scattering rate $N_\mathrm{mc}\Gamma$ being larger than the decoherence rate $\gamma$. Hence, there is a threshold atom number. Far below this threshold we expect purely spontaneous single-atom emission. Well above threshold, collective decay dominates. Between these two regimes a smooth transition takes place.
We determine the threshold to the SF regime using the peak photon rate scaling of SF bursts with $N_\textrm{mc}$. In the regime of collective decay the model developed in \cite{SBP24} recovers the characteristic quadratic scaling of superradiance. 
In the regime of spontaneous, independent emission, we expect a linear scaling of the peak amplitude with $N_\textrm{mc}$. We therefore determine the threshold by fitting the maximum of two linear slopes to the peak amplitude vs $N_\textrm{mc}$ data in double-logarithmic plots as shown in Fig.~\ref{fig:example_threshold}. The green linear slope represents collective and the red one independent scaling. The intersection of both yields the SF threshold. The orange line shows the maximum of the two linear slopes.
Note that below the SF threshold we expect deviations from linear scaling with $N_\textrm{mc}$ as independent spontaneous decay is not affected by decoherence which is accounted for in the model of the maximum cooperation number. Therefore any $N_\mathrm{mc}$ scaling below the threshold should not be interpreted physically. Above the threshold we observe quadratic scaling. However the fit's slope in the log-log space is not always 2 as would be expected in the ideal case. This is due to the fact that the underlying parabola often reveals a small offset if fitted and from the interplay of both power scalings at the threshold. 

\section{Additional Discussions}
\subsection{Impact of Stokes Gain}
\label{sec:Stokes-gain}
The additional dephasing in backward direction results in a smaller Stokes compared to the forward direction.  
The Stokes gain can be estimated as $G_s\approx \langle( 2\alpha_0\Gamma_R)/\gamma\rangle\rangle_r$ \cite{SBP24} where $\langle ...\rangle_r$ denotes a radial average. Here $\alpha_0$ is the peak optical density at a given $N$ and $\gamma$ the total effective decoherence rate of the ensemble.
However the total dephasing is still dominated by the spatial inhomogeneities described in \cite{SBP24} resulting in a relatively small impact of Doppler broadening. The resulting difference in Stokes gain of up to $15\%$ yields a difference of only $7\%$ in the directionality $\kappa$ and does not explain all the differences between TWA and experiment.


\subsection{Comparison of Directionality for TWA Simulations and Experiment}
\begin{figure}[h]
     \includegraphics[width=\columnwidth]{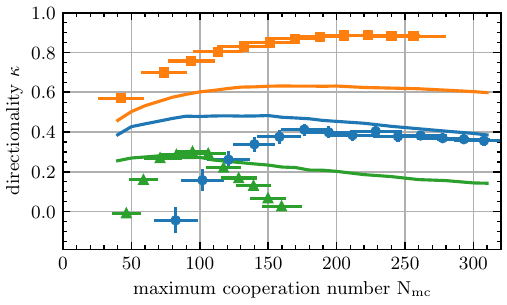}
     \caption{
     	\label{fig:Exp_vs_TWA}
        Overlay of data obtained by the full TWA simulations including motion and the experiment corresponding to $\Gamma_R / (2\pi) = 20(2)~$kHz, $33(2)$~kHz, $67(2)~$kHz. This results in $\sigma_\mathrm{v}$ of 5.0(2)(\textcolor{orange}{\scriptsize$\blacksquare$}), 3.0(1)(\textcolor{RoyalBlue}{$\bullet$}) and 1.50(2)(\textcolor{OliveGreen}{$\blacktriangle$}), respectively.}
\end{figure}

Even though we do not claim quantitative agreement between our simple models and the experiment we would like to provide a direct comparison of the dynamic TWA and experimental results for the interested reader (see Fig.~\ref{fig:Exp_vs_TWA}). As pointed out in the manuscript, no quantitative agreement can be observed or expected due to the underlying strong simplifications of our model. These include the mapping onto smaller emitter numbers underestimating spontaneous processes, the neglect of Stokes gain, and the representation of our ensemble by a 1D model.

\subsection{Cross-Correlation Between Forward and Backward Modes}\label{sec:app_CCF}
\begin{figure}[b]
     \includegraphics[width=0.9\columnwidth]{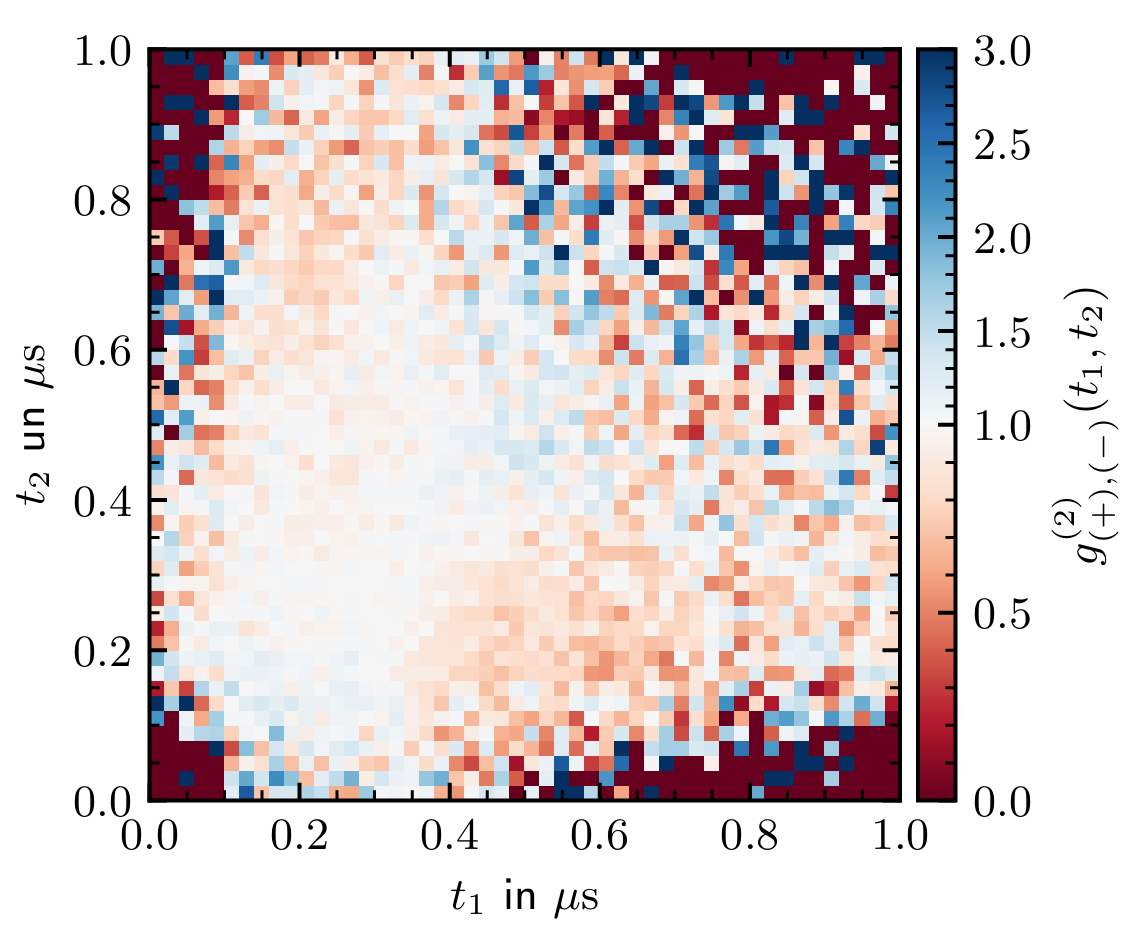}
     \caption{
         \label{fig:CCFs}
     Measured two-time cross-correlation in the $(\pm)$ directions for $N_\mathrm{mc} = 260$, $P_P\approx75$~nW. Note that the colorbar scale is different from Fig. 2.}
\end{figure}

In addition to the auto-correlation measurements presented in the main text, we here briefly discuss the cross-corelation between the (+) and (-) emission channels. These measurements were obtained using protocol A to verify the simultaneous occurrence of bursts in both directions (see section II B). During these measurements we did not determine $\sigma_\mathrm{v}$. 
Exemplary results are shown in Fig.~\ref{fig:CCFs} for low directionality. As we can see, the cross-correlation is close to unity during the bursts ($t_{1,2} \lesssim 0.6~$µs) indicating uncorrelated spontaneous events triggering emission into opposite directions.
Our explanation is as follows:
The fields emitted in each direction, project the collective spin state onto two distinct spatial phase patterns: $(k_p-k_0)z_j$ for (+) and $(k_p+k_0)z_j$ for (-) direction. For an infinitely extended disordered ensemble with random atomic positions, these two patterns are orthogonal; the cross-term $\sum_{j\neq l}e^{i2k_0 (z_j-z_l)}  \langle\sigma_j^+ \sigma_l^-\rangle$ averages to zero.
Physically, a spontaneous photon emitted into (+) imprints a coherence with the spatial structure $(k_p-k_0 ) z_{jl}$  onto the ensemble. This coherence drives stimulated emission into (+) but has no overlap with the (-) mode, whose spatial filter $(k_p+k_0) z_{jl}$  differs by the rapidly oscillating factor $e^{i2k_0 z_{jl} }$. The collective avalanche in each direction is therefore seeded and amplified independently, consistent with the observed absence of cross-correlations. 
We note that the simulations show a similar behavior. However, as the exact pump power was not determined during this measurement, quantitative comparison to the experimental data is currently not possible and would require further measurements planned in the future. 


%

\end{document}